\documentclass[epj,nopacs]{svjour}

\usepackage{multicol}
\usepackage{multirow}
\usepackage[Large]{subfigure}
\usepackage{xcolor}
\usepackage{soul}
\usepackage{stfloats}
\usepackage{float}
\usepackage{hyperref}
\hypersetup{hidelinks,colorlinks=true,allcolors=blue,pdfstartview=Fit,breaklinks=true}
\soulregister\cite7 
\soulregister\ref7
\sethlcolor{yellow}
\setulcolor{red}
\usepackage{amsmath}
\usepackage{siunitx}
\usepackage{amssymb}
\usepackage[final]{changes}
\definechangesauthor[name={libo},color=blue]{libo}
\definechangesauthor[name={li gang},color=red]{lig}

\usepackage[T1]{fontenc} 

\usepackage[switch]{lineno}

\begin{document}
\title{Performance studies of jet flavor tagging and measurement of $R_b(R_c)$ using ParticleNet at CEPC}
\author{Libo Liao\inst{1,}\thanks{\emph{liaolibo@ihep.ac.cn}}, Shudong Wang\inst{2,4}, Weimin Song\inst{3}, Zhaoling Zhang\inst{3},  \and Gang Li\inst{2,}\thanks{\emph{li.gang@ihep.ac.cn} (corresponding author)}
}                     
%
%
\institute{Guangxi Key Laboratory of Machine Vision and Intelligent Control, Wuzhou University, 82 Fumin Third Road, Wanxiu District, Wuzhou, China \and Institute of High Energy Physics, Chinese Academy of Sciences,  19B Yuquan Road, Shijingshan District, Beijing, China \and College of Physics, Jilin University, 2699 Qianjin Street, Changchun, China \and University of Chinese Academy of Sciences, 19A Yuquan Road, Shijingshan District, Beijing, China}
\date{Received: date / Revised version: date}
%
\abstract{
Jet flavor tagging plays a crucial role in the measurement of relative partial decay widths of $Z$ boson, denoted as $R_b$($R_c$),  which is considered as a fundamental test of the Standard Model and sensitive probe to new physics. In this study, a Deep Learning algorithm, ParticleNet, is employed to enhance the performance of jet flavor tagging. The combined efficiency and purity of $c$-tagging is improved by more than 50\% compared to the Circular Electron Positron Collider (CEPC) baseline software. In order to measure $R_b$($R_c$) with this new flavor tagging approach, we have adopted the double-tagging method. The precision of $R_b$($R_c$) is improved significantly, in particular to $R_c$, which has seen a reduction in statistical uncertainty by 40\%.  
} 
\titlerunning{Performance studies of jet flavor tagging and measurement of $R_b(R_c)$ using ParticleNet at CEPC}
\authorrunning{Libo Liao et al.}
\maketitle
\section{Introduction}
\label{intro}

The measurement of the relative partial decay widths of $Z$ boson, $R_q = \Gamma_{q\bar{q}}/\Gamma_h$, {where $\Gamma_{q\bar{q}}$ and $\Gamma_h$ are the partial decay width of $Z\to q\bar{q}$ and the total hadronic decay width respectively}, plays a crucial role in testing the Standard Model (SM)~\cite{Glashow,Weinberg} and searching for new physics. Particularly, $R_b$ is {sensitive to the loop corrections to the $Zb\bar{b}$ vertex, potentially sensitive to new physics contributions}~\cite{SUSY}. The decay width to a quark-antiquark final state can be expressed as~\cite{Vysotsky:1996he}
\begin{eqnarray}
\Gamma(Z\to q\bar{q}) & =&  \frac{G_F M_Z^3}{2\sqrt{2}\pi} (g^2_{Aq} R_{Aq} + g^2_{Vq} R_{Vq})~,
\end{eqnarray}
where $g_{Aq}$ and $g_{Vq}$ are the axial and vector coupling constants, respectively, and $R_{Aq}$ and $R_{Vq}$ are radiation factors to account for the final state Quantum Electrodynamics (QED) and Quantum Chromodynamics (QCD) corrections.
The electroweak radiative corrections to the propagator and the $Zq\bar{q}$ vertex are effectively accounted for in the $g_A$ and $g_V$ couplings. The QED and QCD corrections at first order are flavour blind and can be represented as 
\begin{equation}
    R_{Aq}\approx R_{Vq}\approx 1+\frac{\alpha_s(M_Z)}{\pi}~,
\end{equation}
so that the counterparts of the denominator and numerator cancel each other out in the ratio $\Gamma_{q\bar{q}}/\Gamma_h$. 

The latest world averages of $R_b$ and $R_c$, which are dominated by the measurements of experiments on the LEP and the {SLC}~\cite{L3:1999aer,OPAL:1998kxc,DELPHI:1998cnd,ALEPH:1997xqy,SLD:2005zyw,ALEPH:2005ab}, and {the combination results} of  Gfitter Group~\cite{Gfitter2018} for $R_b$ and $R_c$ are shown in Table~\ref{tab:rbrcvalue}. 
It is apparent that the theoretical uncertainties given by the Gfitter Group are smaller than the experimental results by about two orders of magnitude. Therefore, It is a promising way to search new physics beyond the Standard Model by reducing the uncertainties in both experimental and theoretical domains.
\begin{table}[htbp]
\centering
    \caption{{$R_b$ and $R_c$ values in experiment and Gfitter.}}
    \label{tab:rbrcvalue}
    \begin{tabular}{ccc}
    \hline
    \hline
         & Experiment & Gfitter results  \\
    \hline
         $R_b$&$0.21629\pm 0.00066$ & $0.21582\pm0.00011$\\
         $R_c$&$ 0.1721\pm 0.0030$ & $0.17224\pm0.00008 $\\
    \hline
    \hline
    \end{tabular}
\end{table}

Various approaches have been used to measure the $R_b$($R_c$), such as double tagging, multi-tagging, etc. However, the precision was limited by the statistics and detector performance. Recently, a few electron-positron colliders, such as the CEPC~\cite{CDR-D} and the FCC-ee~\cite{FCC:2018evy}, were proposed to perform precision Higgs and electroweak studies. These facilities are going to deliver huge statistics of data at $Z$ pole, $W$ threshold, and about 240 GeV to maximize the production cross section of Higgs-struhlung process, and so on. 
It is natural that these experiments will adopt both new detector and software technologies to achieve the best performance in the detection and reconstruction of physics objects, especially for jets.  

To measure $R_b$($R_c$), jets are essential physics objects. Therefore, good jet reconstruction algorithms are key ingredients, in particular, jet flavor tagging. Jets from different quarks have different characteristics. For instance, {the final states of $b$-jets} usually have a wider energy distribution, and the vertex displacement of tracks in a $b$-jet are larger than those of other jets because of the long lifetimes of {$b$-flavored hadrons}, and {so on}. 

The LCFIPlus~\cite{LCFIPlus} based on {the TMVA package~\cite{TMVA2007}}, is used for the International Linear Collider (ILC)~\cite{ilc1,ilc2}, the CEPC,  and the FCC-ee physics performance study and detector optimization.  The CEPC delivers great $b/c$-tagging performance thanks to its high precision vertex detector. The $b$-jets can be tagged with an efﬁciency of 80\% at a purity of 90\%. Compared with $b$-tagging, $c$-jet tagging is particularly challenging as charm hadrons have relatively shorter lifetimes than bottom ones and suffer more backgrounds. Therefore, an efficiency of 60\% and a purity of 60\% can be achieved for the $c$-jet tagging.
The FCC-ee also investigated jet flavor tagging by developing its own deep learning flavor tagging tool, ParticleNetIDEA~\cite{Bedeschi:2022rnj,Gautam:2022szi}. It is suggested that this methodology yields performance that is commensurate with those reported in the present study.
In this article, the performance of {jet} flavor tagging of the CEPC baseline detector is improved using a new deep learning (DL) algorithm, ParticleNet~\cite{Qu:2019gqs}. In addition, another novel DL algorithm, Particle Flow Network (PFN)~\cite{Komiske:2018cqr}, is used for comparison and cross-checking. 
The {article is organized as follows.} The simulation, reconstruction software, and Monte Carlo (MC) samples are introduced in Section~\ref{sec:detector}; the DL algorithms and the results of jet  flavor tagging are presented in Section~\ref{sec:ft}; then the measurement of $R_b$($R_c$) is discussed in Section~\ref{sec:rb};  and a summary is given in Section~\ref{sec:conclusion}. 

\section{Detector, software, and samples}
\label{sec:detector}
\begin{figure}[ht]
 \resizebox{0.45\textwidth}{!}{{\includegraphics{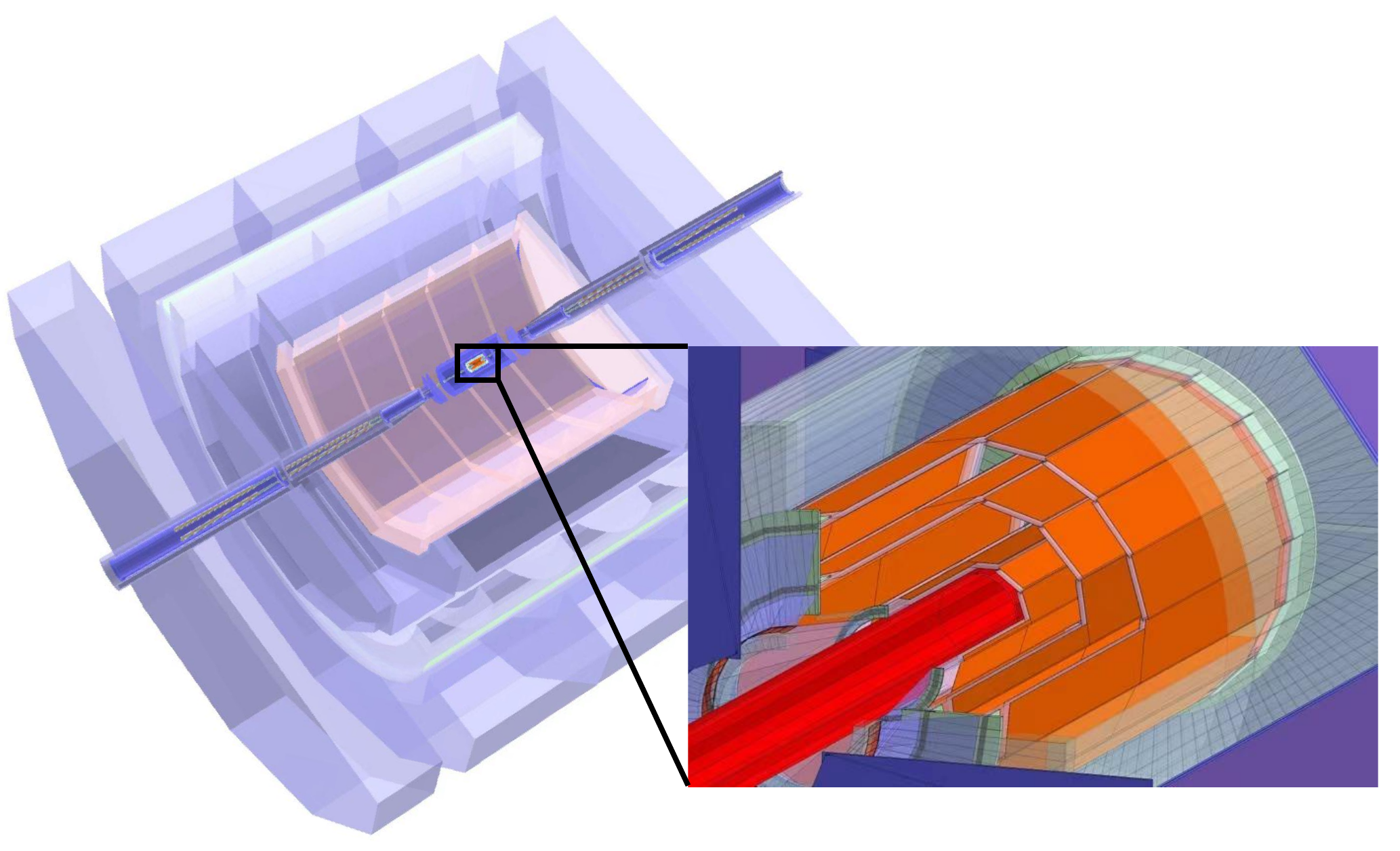}}}
    \caption{ The CEPC baseline detector. 
    The left is the $r-\phi$ view of the detector. In the barrel from inside to outside, the detector is composed of a silicon pixel vertex detector, a silicon inner tracker, a TPC, a silicon external tracker, an ECAL, an HCAL, a solenoid of 3 Tesla, and a muon detector. The right is the silicon pixel vertex detector, which consists of 3 concentric cylindrical double-layers of high spatial resolution.}  
    \label{fig:cepcbaselinedetector}
\end{figure}

The study is based on the CEPC baseline detector, which is advanced {design} from the International Large Detector~\cite{ild} on the ILC and optimized to meet the physics requirement of the CEPC, as shown in Fig.~\ref{fig:cepcbaselinedetector}. The baseline detector is designed according to the Particle Flow Algorithm~\cite{Arbor}, which could reap a better precision and efficiency of reconstructed objects by using the most suitable sub-detectors. From the inside out, the detector includes a silicon pixel vertex detector, a silicon tracker, a time projection chamber (TPC), a calorimetry system which includes an electromagnetic calorimeter (ECAL) and a hadronic calorimeter (HCAL) of very high granularity, and a muon detector embedded inside the return yoke of a solenoid magnet system which provides a magnetic field of 3 Tesla.

The vertex detector consists of six layers of silicon pixel sensors {at radii between 1.6 and 6.0 cm }with excessive spatial resolution {of $\sim 5~\mu\text{m}$}. 
The resolution in $r\phi$ plane can be parameterized by 
\begin{equation}
    \sigma_{r\phi} = a \oplus \frac{b}{p\text{(GeV)}\sin^{3/2}\theta},
\end{equation}
where $\sigma_{r\phi}$ denotes the impact parameters resolution, $p$ is the track momentum, and $\theta$ is the polar track angle, $a=5~\mu \text{m}$ and $b=10~\mu\text{m}\cdot\text{GeV}$.
The silicon tracker is made of 4 components, which are the Silicon Inner Tracker, the Silicon External Tracker, the Forward Tracking Detector, and the End-cap Tracking Detector. The Time Projection Chamber is designed within the framework of the LCTPC collaboration~\cite{LCTPC} and provided a large number of hits to enhance track finding efficiency. The ECAL and HCAL are each composed of 1 barrel and 2 end-cap sections. The detailed description of the CEPC baseline detector is in Ref.~\cite{CDR-D}.

The MC samples for this study are produced with the CEPC full simulation, reconstruction, and analysis framework~\cite{cepcsoft}. The physics processes are generated with {\sc WHIZARD} 1.9.5~\cite{whizard}. PYTHIA 6~\cite{pythia6} is then used for hadronization. MokkaPlus~\cite{MokkaPlus},  a GEANT4-based~\cite{geant4} detector simulation tool, is used to model the detector response. Arbor~\cite{Arbor} is used to reconstruct physics objects including tracks, photons, and neutral hadrons, and LCFIPlus~\cite{LCFIPlus} is used to reconstruct (secondary) vertices and jets. 

There are 3 hadronic decay modes of $Z$ boson used for jet flavor tagging in this study, which are $e^+e^- \to Z \to b\bar{b}$, $c\bar{c}$,  and $q\bar{q}(u\bar{u}/d\bar{d}/s\bar{s})$. 
For each process, 450,000 events are produced, which has 900,000 jets in total. The jets are reconstructed using $e^+e^-$-$k_t$ algorithm in LCFIPlus~\cite{LCFIPlus}, where all particles, including the reconstructed primary and second vertices,  are clustered into two jets.  

\section{Jet flavor tagging with ParticleNet}
\label{sec:ft}
In this study, ParticleNet is utilized as the nominal algorithm, while PFN is employed as a comparison and cross-check method. 

Based on the particle cloud representation, which treats a jet as an unordered group of particles, an effective algorithm, ParticleNet, has been developed. It is a customized neural network model using Dynamic Graph Convolutional Neural Network (DGCNN)~\cite{DGCNN} for jet tagging. 

ParticleNet has several advantages. First, it can deal with the varying number of particles in an event, which is common in experimental high energy physics. Second, the algorithm is designed to respect the particle permutation invariance, {which refers to the fact that the algorithm does not assume any special order of the particles in a jet.} Third, ParticleNet makes extensive use of EdgeConv~\cite{DGCNN} operations to update the graph representation dynamically. The study in Ref.~\cite{DGCNN} shows that it is beneficial to recompute the graph using nearest neighbors in the feature space produced by each layer. With dynamic graph updates, the jet (sub-)structure can be probed hierarchically, which leads to better performance than keeping the graph static.  Last but not least, ParticleNet could exploit local neighborhood information explicitly while most of the other DL algorithms could only use global symmetric features. 
 
\subsection{Visualizing the data sets}

\begin{figure*}
\centering
    \resizebox{0.9\textwidth}{!}{
    \subfigure[]{\includegraphics{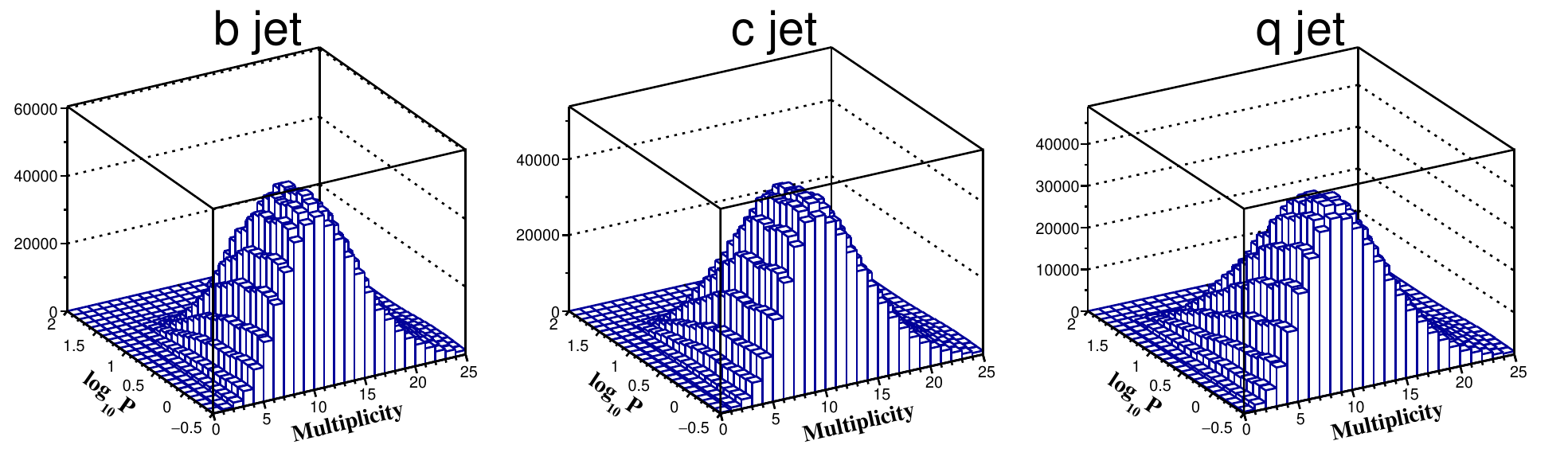}\label{fig:2Dlego}}}
     \resizebox{0.9\textwidth}{!}{
    \subfigure[]{\includegraphics{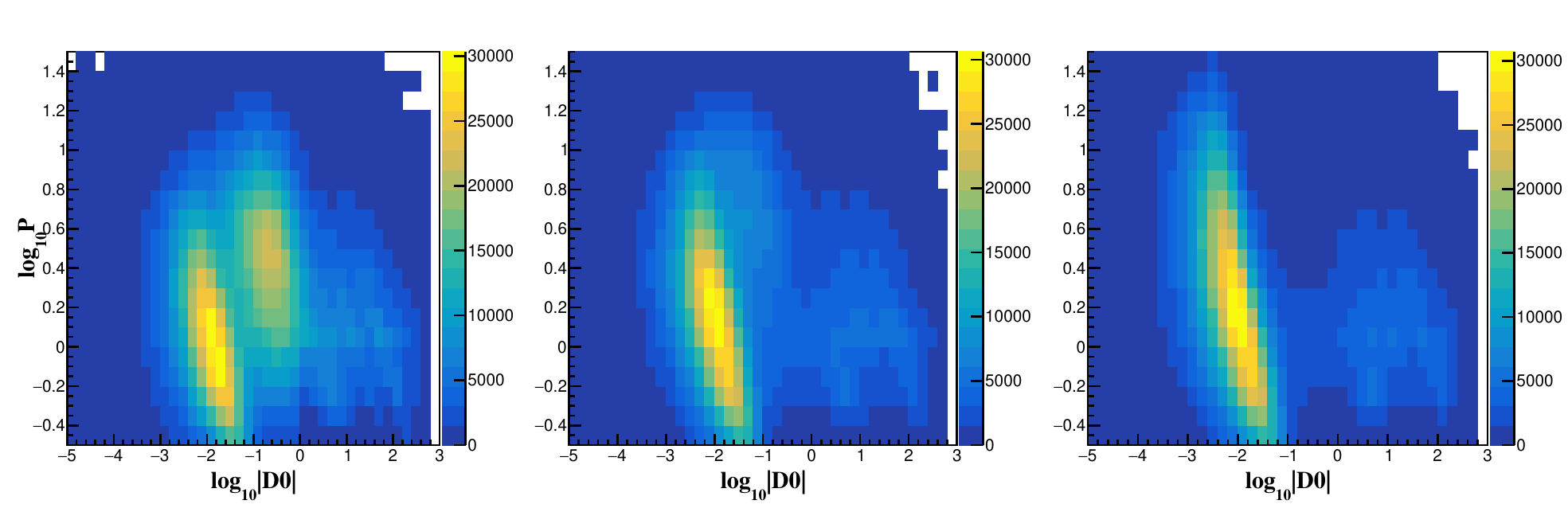}\label{fig:2DMD}}}
     \resizebox{0.9\textwidth}{!}{
     \subfigure[]{\includegraphics{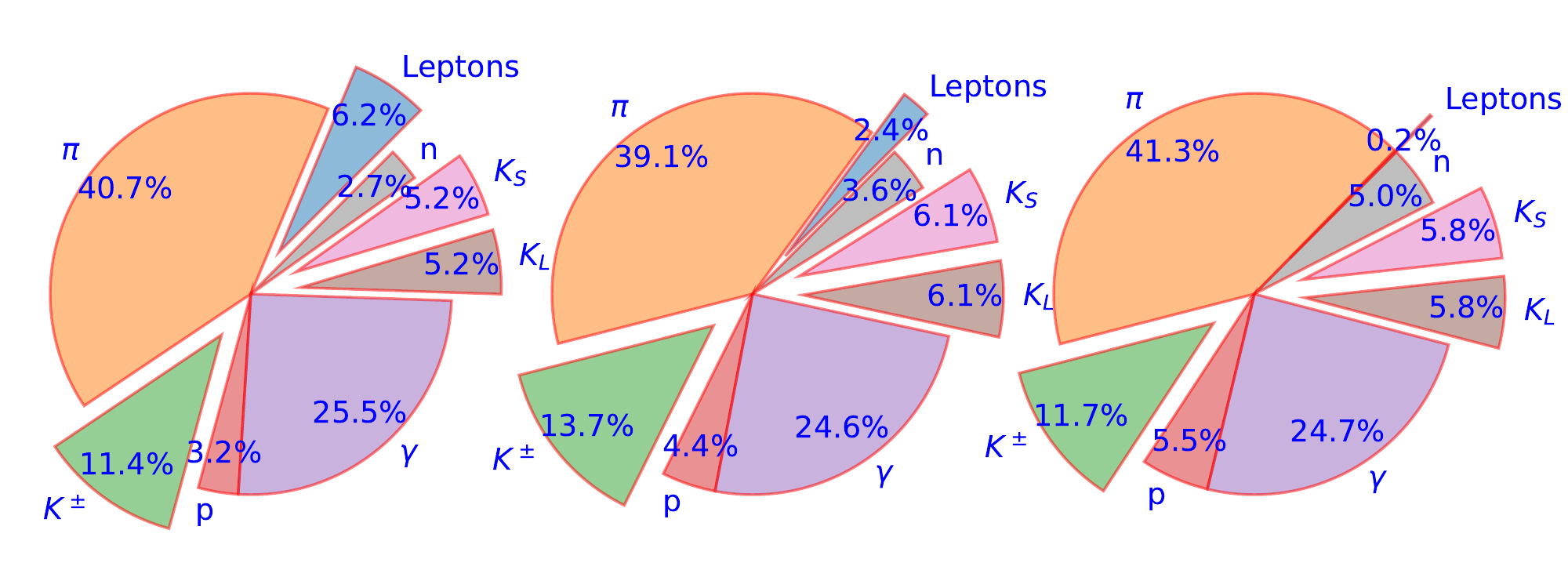}\label{fig:quarkMom}}}
    \caption{The feature plots of $b, c,$ and $q$ in jet level. The 2-dimensional diagram of charged multiplicity versus momentum distribution is shown in the top panel; the 2-dimensional diagrams of momentum versus D0 are shown in the middle panel;  the fractions of all particle types in $b\bar{b}$, $c\bar{c}$, and $q\bar{q}$ weighted by momentum in the bottom panel, where the PID is based on the MC truth.}
    \label{fig:discrepancy of bclt}
\end{figure*}

The jet flavor tagging algorithm is based on features of the data sets. In this study, these features could be categorized into  three types. The first type is related to jet kinematics, such as multiplicity, momentum distribution, etc. The second is the impact parameters of the charged tracks, which are very informative for $b$-tagging. The last one is the types of particles in a jet, i.e., particle identification (PID). 
Those could be expected that the multiplicity of $b$-jet should be larger than the others because of higher masses of $b$-flavored hadrons and that the tracks should have larger vertex displacement because of their longer lifetimes, etc. 
Considering three types of jets to be studied, some distributions are shown in Fig.~\ref{fig:discrepancy of bclt}. Figure~\ref{fig:2Dlego} is the multiplicity versus momenta of tracks, where it can be seen that the number of tracks in $b$ jets is slightly larger than those of $c$ jets and light ($q$) jets, which is consistent with the decay properties of heavier $B$ hadrons. The distribution of impact parameters  versus the momenta is shown in Fig.~\ref{fig:2DMD}. Clear patterns can be observed: $b$ jets have the significant contribution of larger impact parameters and of energetic tracks compared with $c$ and $q$ jets. Figure~\ref{fig:quarkMom} shows the momentum weighted fractions of different particle types in the three physics processes. It is clear that $b$ quarks produce more energetic leptons, $c$ quarks produce slightly more energetic kaons. 
All the above are consistent with our expectations. 

In this data-set, kinematic information, i.e., the
four momentum ($p_x$,$p_y$,$p_z$,$E$), and vertex displacements of each particle when available, are listed in Tab.~\ref{tab:inputvariables}. The ($\cos\theta$, $\phi\sin\theta$) are used as coordinates~\cite{Li:2020vav} to compute the distances
between particles in the first EdgeConv block. They are
also used together with some other variables,  such as $\Delta R$, PID, $E$, $Q$, $\log E$, $\log P$, $D_0$, $Z_0$, $D_0/\sigma_{D_0}$, $Z_0/\sigma_{Z_0}$, and the prob which is defined as 
\begin{equation}
    \text{prob} = \int_{\chi^2}^\infty p(x,N)dx,
\end{equation}
where $\chi^2 = (D_0/\sigma_{D_0})^2+(Z_0/\sigma_{Z_0})^2$,  $p(x,N)$ is the probability density function of the chi-square distribution, and $N ( =2)$ is the number of degrees of freedom.

\begin{table*}[thbp]
    \centering
    \caption{Variables used in the DL algorithms. }
    \begin{tabular}{cl}
    \hline
    \hline
        Variable & Definition \\
    \hline
        $\cos\theta$ & cosine of polar angle of particle \\
        $\phi\sin\theta$ & azimuth angle times sine of polar theta of particle \\
        \hline
        $\Delta R$ & $\sqrt{ \delta\theta^2  + \delta \phi^2}$, angular separation between the particle and the jet axis \\
        PID & particle ID \\
        $E$ & energy of a particle \\
        $Q$ & electric charge of a particle \\
        $\log E$ & logarithm of the particle's energy\\
        $\log P$ &logarithm of the particle's momentum\\
        $D_0$ & impact parameter of a track in the r-$\phi$ plane  \\
        $Z_0$ & impact parameter of a track along the $z$ axis \\
        $D_0/\sigma_{D_0}$ & significance of the impact parameter in the r-$\phi$ plane \\
        $Z_0/\sigma_{Z_0}$ & significance of the impact parameter along  the $z$ axis  \\
        prob & the probability for a certain Chi-squared and number of degrees of freedom\\
    \hline
    \hline
    \end{tabular}
    \label{tab:inputvariables}
\end{table*}

\subsection{Deep learning algorithms and configuration}

The ParticleNet used in this paper consists of three EdgeConv blocks, a global average pooling layer, and two fully connected layers.  
The number of channels $C$ for each EdgeConv block is (64, 64,64), (128, 128, 128), and (256, 256, 256), respectively. After the EdgeConv blocks, a channel-wise~\cite{channelwise} global average pooling operation is applied to aggregate the learned features over all particles in the cloud. This is followed by a fully connected layer with 256 {neurons} and the ReLU activation~\cite{relu}. A dropout layer~\cite{dropout} with a drop probability of 0.1 is included to prevent overfitting. A fully connected layer with $N$ neurons, followed by a softmax function, is used to generate the output, where the $N$ is the number of categories in a classification task. For the number of nearest neighbors $k$ for all three blocks, some optimization is performed, which shows that 12 for jet tagging is optimal. The configuration of PFN is directly taken from the Ref.~\cite{Komiske:2018cqr}, since it is only used for cross-checking.

\subsection{Training and evaluation}

Both ParticleNet and PFN are implemented and running with 8 Intel$^\circledR$ Xeon$^\circledR$ Gold 6240 CPU cores and 4 NVIDIA$^\circledR$ Tesla$^\circledR$ V100-SXM2-32GB GPU cards at the IHEP GPU farm. During training, the common properties of the neural networks include a categorical cross-entropy loss function, the Adam optimization algorithm~\cite{adam}, a batch size of 1,024, and a starting learning rate of 0.005. 900,000 jets are used for each process, therefore, the total number of jets is 2,700,000. The full data-set is split into training, validation, and test samples according to the fraction of 7:1.5:1.5. The monitoring of loss and accuracy on training and validation shows that the algorithm converges well and there is no obvious over-training.

The computation consumption of ParticleNet and PFN algorithms could be estimated. Only the total consumption of GPU and CPU is used for estimation since all the computing resources could only be accessed indirectly via a workload manager server.  ParticleNet takes about 190 minutes for training (30 epochs) and 3 minutes for inference. PFN takes about 30 minutes for training (80 epochs) and less than a minute for inference. Both two methods could be finished on a reasonable time scale. However, it could be a problem for the study of $10^{12}~Z$ bosons and solved by hardware development in the next decades.

\subsection{Performance}
 
Both ParticleNet and PFN outperform the LCFIPlus in terms of jet flavor tagging. 
The accuracy of two novel algorithms, which is defined as the fraction of correctly classified jets, are summarized in Tab.~\ref{tab:acc}, together with those in Ref.~\cite{yangfan}. 
PartcileNet could achieve an accuracy of about 87.6\%, which is at least 9\% better than those in Ref.~\cite{yangfan}.

\begin{table*}[ht]
\centering
	\caption{The accuracy of different algorithms for jet flavor tagging. In this study, ParticleNet is trained 9 times using randomly initialized weights, and the results from the median-accuracy are shown, while PFN is trained only once and the uncertainty from randomly initialized weights is negligible.}
	\label{tab:acc}
	\begin{tabular}{cccccccc}
	\hline
	    \hline
	    Algorithm & ParticleNet & PFN & DNN & BDT & GBDT & gcforest & XGBoost \\
		\hline
		Accuracy  & 0.876 & 0.850 & 0.788& 0.776 & 0.794 & 0.785 & 0.801 \\
		\hline
		\hline
	\end{tabular}

\end{table*}

\begin{figure}
    \centering
    \resizebox{0.45\textwidth}{!}{
    \includegraphics{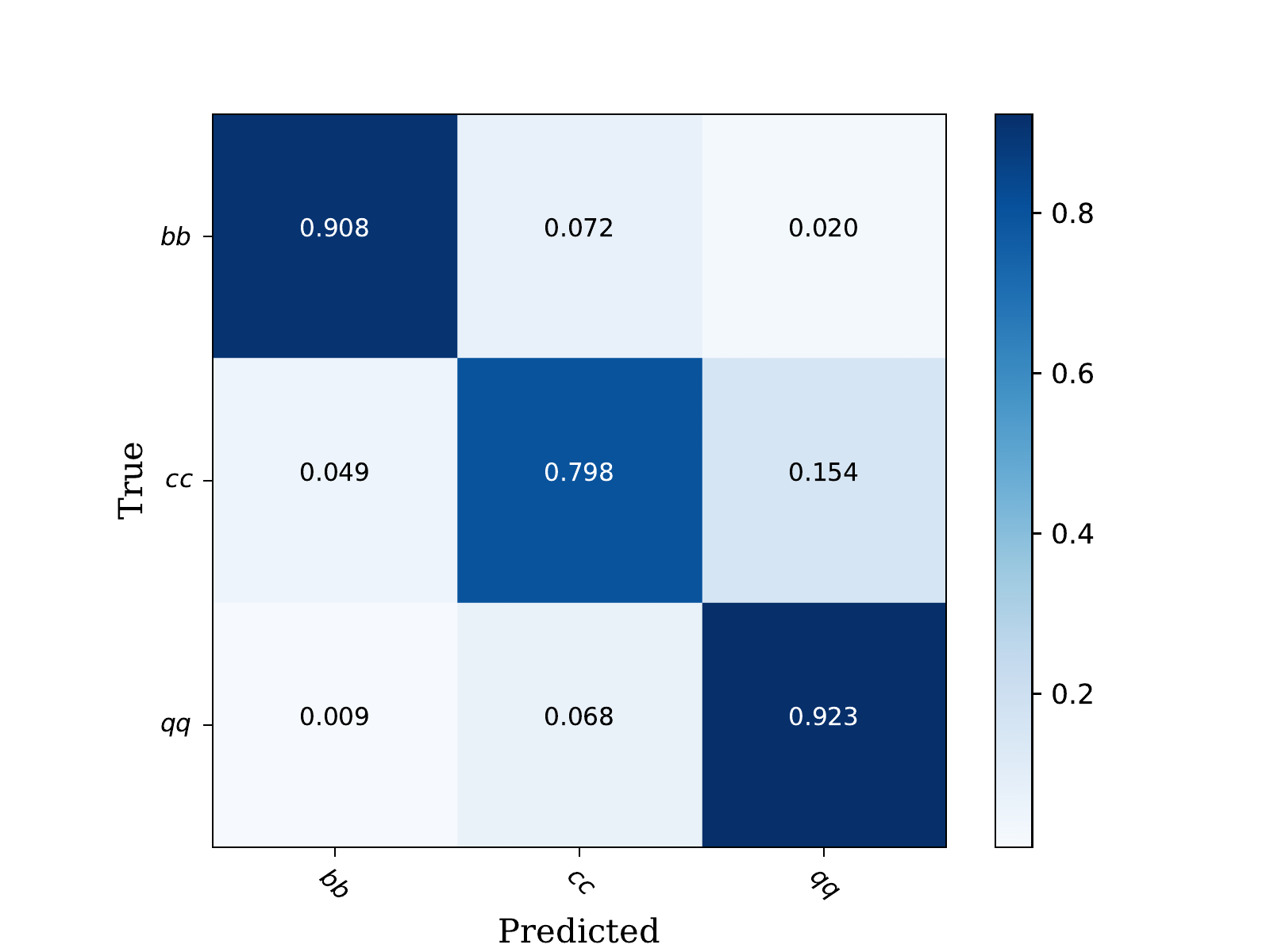}}
    \caption{The confusion matrix with ParticleNet. The training is repeated 9 times using randomly initialized weights, and the results of the training with median-accuracy are adopted.}
    \label{fig:cmrpa}
\end{figure}

The numerical results of efficiencies  and Area Under Curve (AUC) of both algorithms in different jet flavor tagging are listed in Tab.~\ref{tab:jetPNPFN}. The efficiencies, also called recalls, are determined by taking the largest score of a jet predicted by the classifiers, which are the same as the corresponding diagonal terms of the confusion matrix {which is shown in Fig.~\ref{fig:cmrpa}}. It can be seen that the performance of $b$-tagging is always better than $c$- and $q$-tagging. The observation is consistent with the results in Ref.~\cite{yangfan}.

\begin{figure*}[ht]
\centering
    \resizebox{0.45\textwidth}{!}{
        \includegraphics{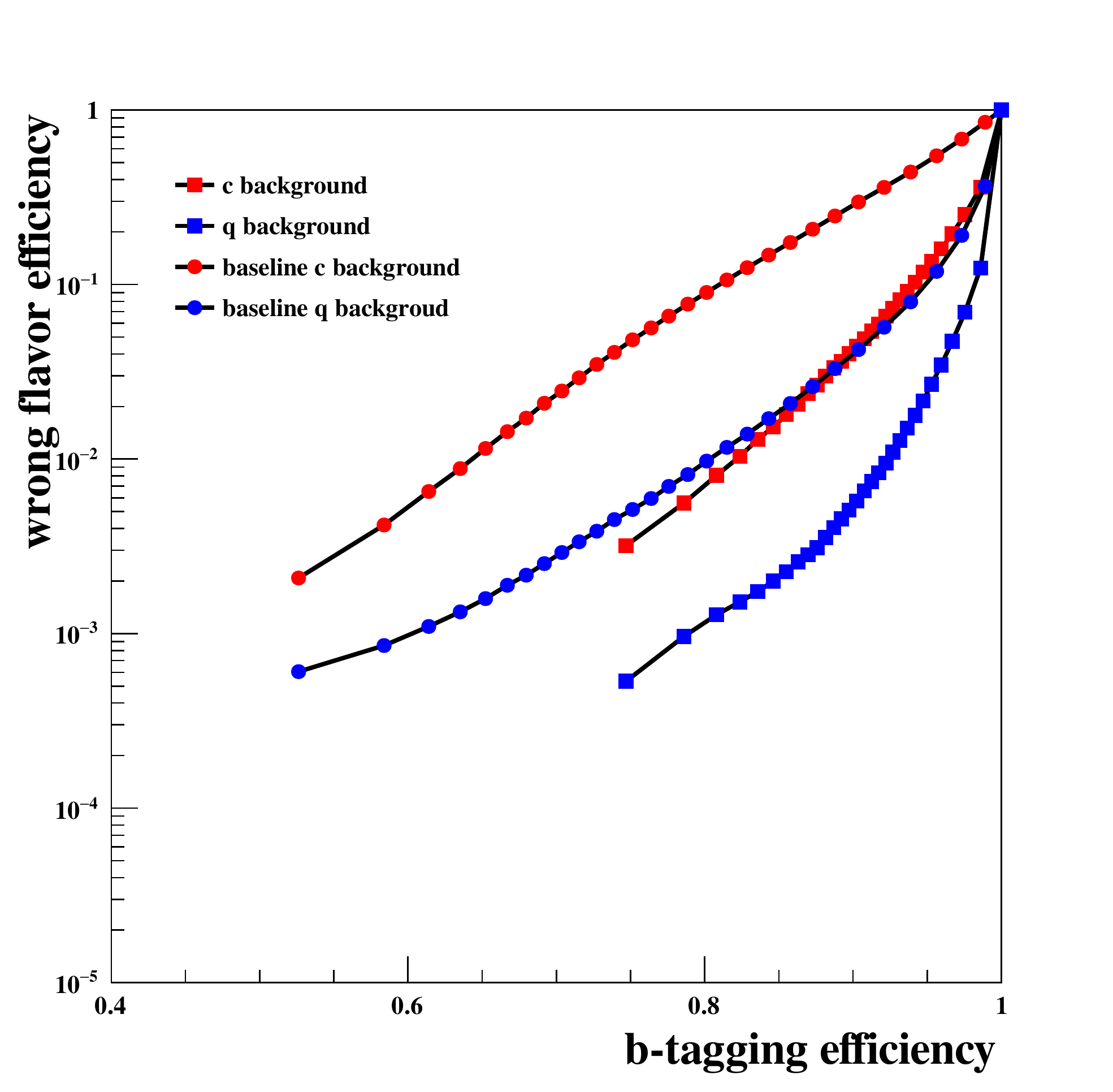}
}
	\resizebox{0.45\textwidth}{!}{
        \includegraphics{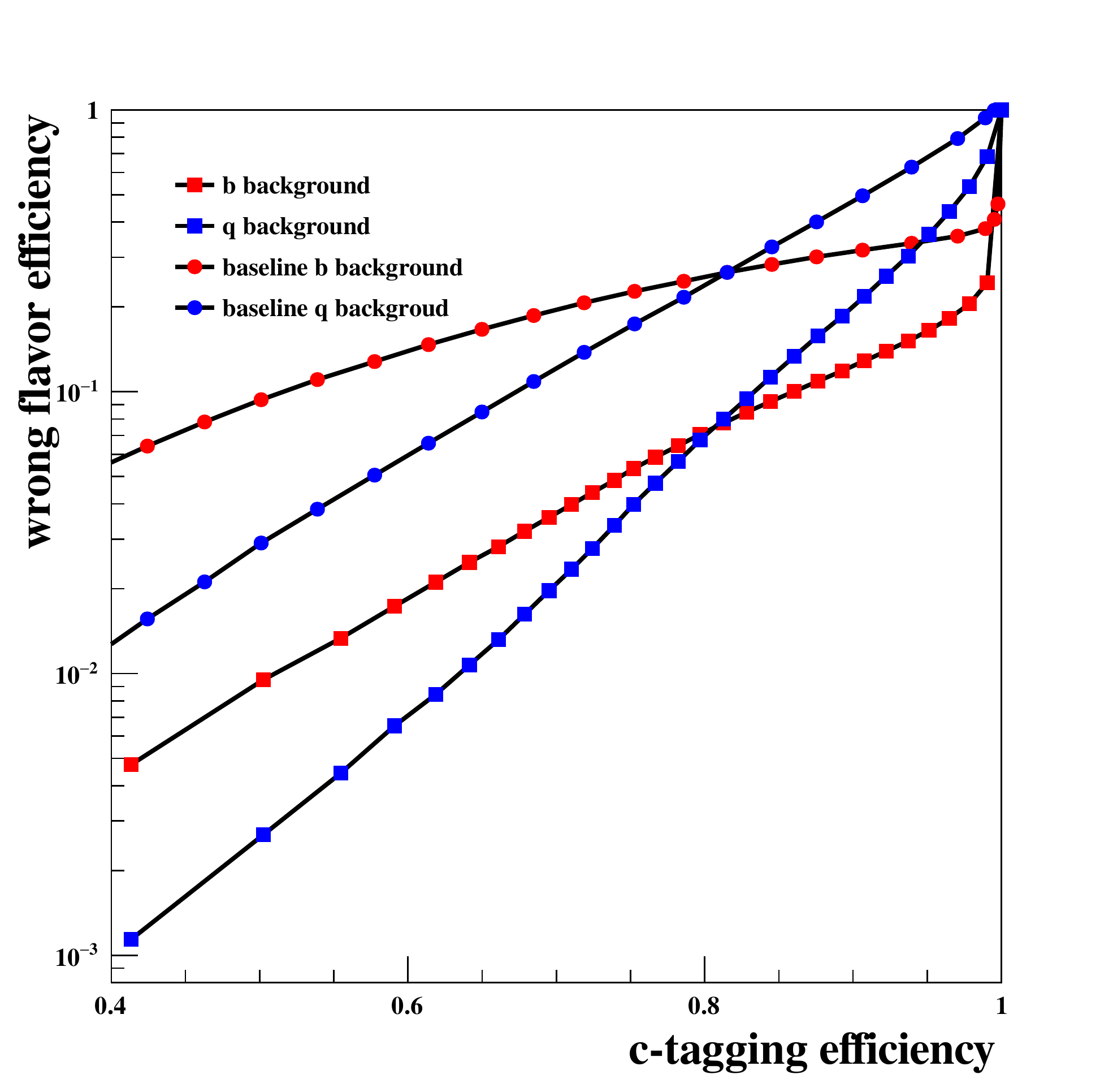}
}
	\caption{Efficiencies for selecting jets with the wrong flavor when tagging $b$ jets (the left panel) and $c$ jets (the right panel). The points and the rectangles are the efficiencies of the CEPC baseline and this study, respectively. The training is repeated 9 times using randomly initialized weights, and the results of the training with median-accuracy are shown.}
	\label{fig:cmrp}
\end{figure*}

The performance of ParticleNet and PFN are generally better than those in Ref.~\cite{yangfan}. This could be explained from two sides, one is that much richer information about a jet is used including four momenta, impact parameters, and PIDs, and the other is that ParticleNet and PFN have a strong inductive bias for representing high energy events. ParticleNet outperforms PFN, which is consistent with the study in Ref.~\cite{Qu:2019gqs}, and the authors explained that "the Deep Sets (PFN) approach does not explicitly exploit the local spatial structure of particle clouds, but only processes the particle clouds in a global way."

\begin{table}[ht]
\centering
    \caption{The performance of ParticleNet and PFN in jet tagging. ParticleNet is trained 9 times using randomly initialized weights and the one with median-accuracy is taken.    }
    \label{tab:jetPNPFN}
    \begin{tabular}{c|cc|cc}
    \hline
    \hline
    \multirow{2}{*}{tag} & \multicolumn{2}{c|}{ParticleNet} & \multicolumn{2}{c}{PFN} \\
         &Efficiency  &AUC& Efficiency & AUC\\
    \hline
    $b$   &0.908& 0.986  & 0.870 &0.979 \\
    $c$   &0.798& 0.951  & 0.765 &0.930 \\ 
    $q$   &0.923& 0.974  & 0.911 &0.966 \\
    \hline
    \hline
    \end{tabular}
\end{table}

An alternative way to show the flavor tagging performance is the tagging efficiencies versus the corresponding wrong flavour efficiencies, as the plots in Fig.~\ref{fig:cmrp}. For $b$-tagging, the main background is from $c$ jets. 
In the case of $c$-tagging, the situation is different. The main background is from light flavour jets at efficiency above 80\%, while it is dominated by misidentified $b$ jets at lower efficiency. 

To demonstrate the physics impacts of jet flavor tagging, a detailed comparison in terms of the product of efficiency and purity, $\epsilon\rho$, is performed. Taking the measurement of $R_b$($R_c$) as an example,  the Eq.~\ref{equ:sig} gives the connection between its statistical uncertainty and $\epsilon\rho$. It is known for decades that maximizing $\epsilon\rho$ is identical to minimizing the statistical uncertainties.
\begin{equation}
    (\Delta R_i)^2 \propto \frac{1}{\epsilon_i\rho_i}.
    \label{equ:sig}
\end{equation}

To compare the performance of various jet flavor tagging methods, some working points are chosen. Table~\ref{tab:XvP} summarizes the numerical results, where LCFIPlus and XGBoost are taken as references. The table shows that the performance of ParticleNet is much better than the others, especially in $c$-tagging. ParticleNet is more than 50\% better compared with LCFIPlus when the efficiency of $c$-taggging is 60\%. A specific example to illustrate the impact is that ParticleNet could improve the statistical uncertainty in counting $c$ jets by 30\% compared with the XGBoost. 
PFN also achieves comparable improvement and confirms the correctness of ParticleNet.

\begin{table}[ht]
\centering
	\caption{The performance of the specific method in different working points, where the results of LCFIPlus are reported in Ref.~\cite{CDR-D}, and the results of XGBoost are reported in Ref.~\cite{yangfan} }
	\label{tab:XvP}
	\begin{tabular}{c|c|cc|cccc}
		\hline
		\hline
		\multirow{2}{*}{tag} & \multirow{2}{*}{$\epsilon_S$(\%)}& \multicolumn{4}{c}{$\epsilon\times\rho$} \\
		\cline{3-6}
			&   & LCFIPlus & XGBoost  & ParticleNet & PFN  \\
		\hline

		\multirow{2}{*}{$b$}   &80&  - &0.747&0.786&0.763\\
		                       &90&0.72&0.713&0.821&0.752\\
		\hline
		\multirow{4}{*}{$c$}   &60&0.36&  -  &0.554&0.485\\
		                       &70& -  &  -  &0.605&0.497\\
		                       &80& -  &0.345&0.597&0.467\\
							   &90& -  &0.292&0.532&0.402\\
		\hline
        \hline
	\end{tabular}
\end{table}

\section{Measurement of relative decay width}
\label{sec:rb}
In the LEP, $R_b$ is measured with various methods, which are based on counting the events with either one or both hemisphere tagged. In this study, jet is akin to hemisphere in LEP and it would be used in the rest of this paper. The observed number of jets of flavor $i$ (single tag), $N_s^{i,\mathrm{obs}}$, and the observed number of jet pairs (double tag), $N_d^{i,\mathrm{obs}}$, are given by: 
\begin{equation}
\begin{split}
    N_s^{i,\mathrm{obs}} 
    =& 2N^{h,\mathrm{pro}}\cdot (R_b\varepsilon_{ib}+R_c\varepsilon_{ic}+R_q\varepsilon_{iq})~,\\
    N_d^{i,\mathrm{obs}} =& N^{h,\mathrm{pro}}\cdot[R_b\varepsilon_{ib}^2(1+C_{ib})+R_c\varepsilon_{ic}^2(1+C_{ic})\\
    &+R_q\varepsilon_{iq}^2(1+C_{iq})]~,
\end{split}\label{eqn:tag}
\end{equation}
where $i(j) = b, c, q$ are flavors of jets,  $C_{ij}$ is the correlation between a jet pair of flavor $j$  when both are tagged as $i$, $\varepsilon_{ij}$ is the efficiency of a $j$ jet being tagged as a $i$ jet, $N^{h,\mathrm{pro}}$ is total number of $Z$ hadronic events produced in collisions, $R_i$ is the relative decay widths of $Z$ to jet pair of $i$.

$R_c$ measurement is more challenging than $R_b$, since the $c$-tagging has less efficiency and less purity than $b$-tagging. Therefore, several methods are employed, such as double tag measurement, charm counting, etc. In fact, the key ingredient of a relative partial width measurement is classifying the signal and background correctly, i.e., jet flavor tagging. 

\begin{figure}[ht]
\centering
    \resizebox{0.4\textwidth}{!}{
    \includegraphics{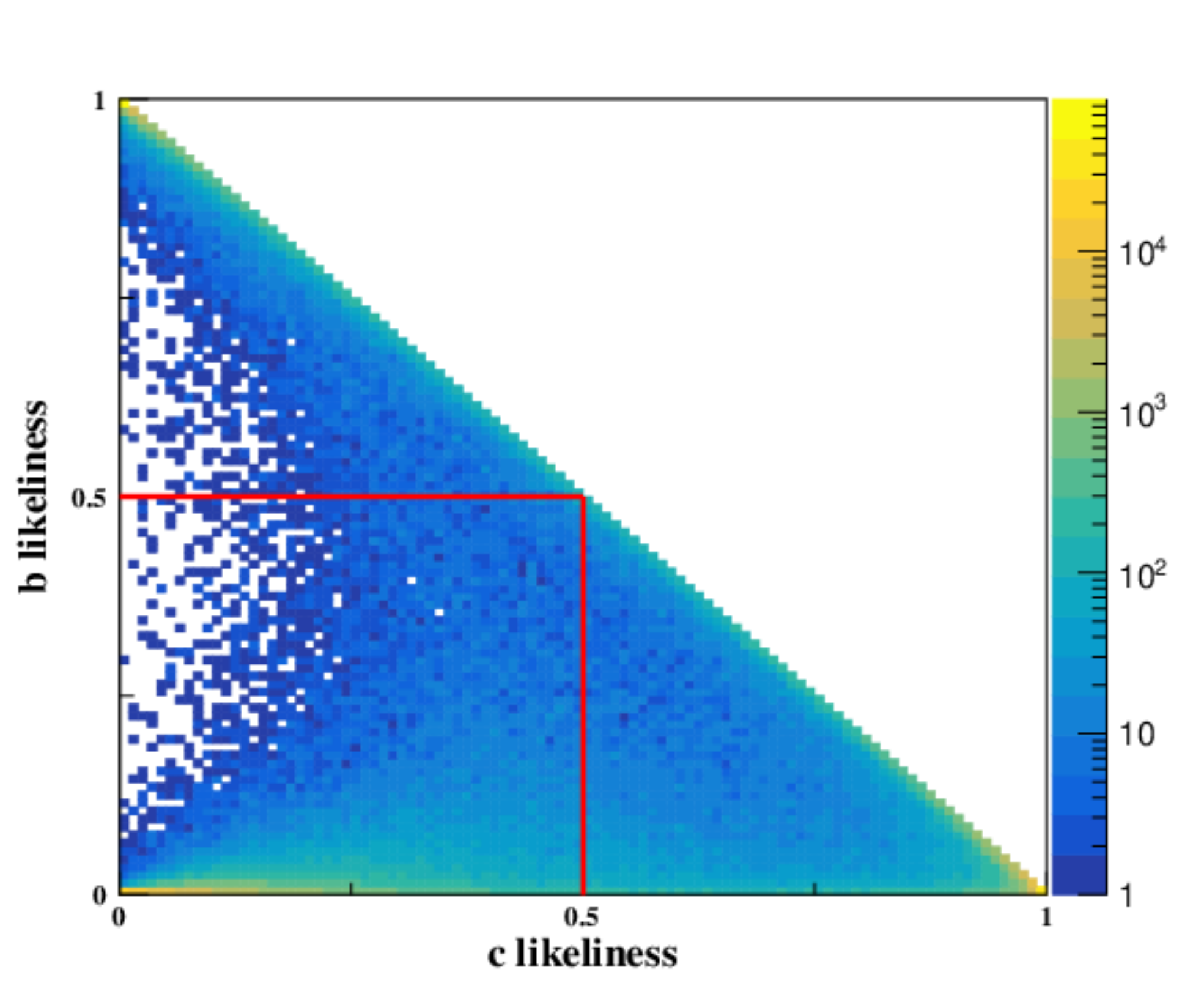}}
    \caption{The two-dimensional distribution of $b$-likeliness versus $c$-likeliness, where the red lines indicate one of the jet classifications. The upper left triangle is the candidate area of $b$, the lower left rectangle is the candidate area of $q$, and the lower right triangle is the candidate area of $c$.}
    \label{fig:Pdistribution}
\end{figure}

To measure $R_b$($R_c$), {the double tag method is deployed, which solves }the Eq.~(\ref{eqn:tag}) to get $R_b$($R_c$) ($R_q = 1- R_b-R_c$ by definition) when a working point is determined. All the $\varepsilon_{ij}$ could be determined by MC simulation and the correlation between jets could be neglected temporarily.
Signal regions of $b$, $c$, and $q$ candidates are defined as the red lines, i.e, working point, in Fig.~\ref{fig:Pdistribution}. 
There are 2 equations for each region, and 6 in total.
As over-determined equations, they could be solved by the least square method. Using the same integrated luminosity assumed in Ref.~\cite{Li:2021zlv}, a toy MC approach is used to calculate the statistical uncertainty of $R_b$($R_c$). A total number of $10^{11}$ $Z$ hadronic decay events is sampled according to Poisson distribution, and then this number is sampled into three categories,  $b\bar{b}$, $c\bar{c}$, and $q\bar{q}$, according to multinomial distribution. The detection and selection procedures are also simulated according to multinomial distribution. Finally, three observed numbers are obtained by adding sampling results.   Now $R_b$($R_c$) could be calculated with the least square method, as well as its statistical uncertainty.  

The results are summarized in Table~\ref{tab:rb}. The measurements of the LEP/SLC~\cite{ALEPH:2005ab}, FCC-ee~\cite{Bernardi:2022hny}, and Ref.~\cite{Li:2021zlv} are also listed for comparison purposes. The uncertainties of relative decay width in LEP/SLC~\cite{ALEPH:2005ab} are primarily limited by statistics.  The template fit~\cite{Li:2021zlv} got excellent precision by using a much larger sample size and more information. The double tag also achieves comparable precision as the template fit on $R_b$, but for $R_c$, the precision is improved by nearly 40\%, thanks to the superior $c$-tagging performance of the DL algorithm compared to LCFIPlus. The statistical uncertainty in FCC-ee is $0.3\times 10^{-6}$. However, it should be noticed that the statistics used at FCC-ee are 50 times larger than those used in Ref.~\cite{Li:2021zlv} and this study. If the same integrated luminosity is assumed, it would be $2.1\times10^{-6}$. So the results of Ref.~\cite{Li:2021zlv}  and this study are much better because the FCC-ee simply extrapolates the results from LEP, while the other two studies employ innovative analysis methods and enhanced detector designs.

\begin{table}[ht]
\centering
	\caption{Statistical uncertainties ($10^{-6}$) of relative decay widths. The results of LEP/SLC~\cite{ALEPH:2005ab}, FCC-ee~\cite{Bernardi:2022hny}, and template fit~\cite{Li:2021zlv} are reported. The flavor tagging methods employed in Template fit and Double tag are also listed.}
	\label{tab:rb}
	\begin{tabular}{ccccc}
		\hline
		\hline
			&	$\sigma_{R_b}$& $\sigma_{R_c} $ &$\sigma_{R_{q}}$&flavor tagging method\\
		\hline
        LEP+SLC       & 659  &3015 & - & -\\
            FCC-ee        & 2.1(0.3)  & -   & - & -\\
		Template fit  & 1.2  & 2.3 & 2.1 & LCFIPlus\\
		Double tag    & 1.3  & 1.4  & - & ParticleNet\\
		\hline
		\hline
	\end{tabular}
\end{table}

\section{Summary and discussion}
\label{sec:conclusion}

This study utilizes two DL algorithms to enhance the performance of jet flavor tagging. ParticleNet, in particular, shows significant improvement in jet flavor tagging, especially with regard to $c$-tagging.  In terms of the product of purity and efficiency, the $c$-tagging is improved by over 50\% compared to the CEPC baseline software when the efficiency of $c$-tagging is 60\%.
It's understandable why ParticleNet achieves significantly better performance.
Compared to the traditional methods, ParticleNet can maximize the usage of information in a jet, as it uses lower-level information, such as momenta, energies, and impact parameters, as input. 
On the other hand, the point-cloud (set) representation, which preserves some important symmetries, has better expressive power for jets~\cite{Lorentz-symmetry}.

$R_b$($R_c$) is used as a test bed to demonstrate the physics impacts of the new DL algorithm. The results indicate that the precision of $R_c$ can be improved by a factor of 1.6 compared to those in Ref.~\cite{Li:2021zlv}. In a high-precision study as this, the systematic uncertainties pose a significant challenge  and require careful investigation in future studies.

\paragraph{Acknowledgements}
This work is partially supported by the National Natural Science Foundation of China (NSFC) (12075271, 12047569) and Guangxi University Young and Middle-aged Teachers Research Basic Research Ability Improvement Project (2023KY0707).

%
%

\end{document}